# Cloudy and Cloud-free Thermal Phase Curves with `PICASO`: Applications to WASP-43b

Nina Robbins-Blanch[1], Tiffany Kataria[2], Natasha E. Batalha[3], and Danica J. Adams[4]
[1] Department of Astronomy and Astrophysics, University of California Santa Cruz, Santa Cruz, CA 95064, USA; nrobbin1@ucsc.edu
[2] Jet Propulsion Laboratory, California Institute of Technology, Pasadena, CA 91109, USA
[3] NASA Ames Research Center, MS 245-3, Moffett Field, CA 94035, USA
[4] Division of Geological and Planetary Sciences, California Institute of Technology, Pasadena, CA 91125, USA

## Abstract

We present new functionality within `PICASO`, a state-of-the-art radiative transfer model for exoplanet and brown dwarf atmospheres, by developing a new pipeline that computes phase-resolved thermal emission (thermal phase curves) from three-dimensional (3D) models. Because `PICASO` is coupled to `Virga`, an open-source cloud code, we are able to produce cloudy phase curves with different sedimentation efficiencies ($f_{sed}$) and cloud condensate species. We present the first application of this new algorithm to hot Jupiter WASP-43b. Previous studies of the thermal emission of WASP-43b from Kataria et al. found good agreement between cloud-free models and dayside thermal emission, but an overestimation of the nightside flux, for which clouds have been suggested as a possible explanation. We use the temperature and vertical wind structure from the cloud-free 3D general circulation models of Kataria et al. and post-process it using `PICASO`, assuming that clouds form and affect the spectra. We compare our models to results from Kataria et al., including Hubble Space Telescope Wide-Field Camera 3 (WFC3) observations of WASP-43b from Stevenson et al. In addition, we compute phase curves for Spitzer at 3.6 and 4.5 $\mu$m and compare them to observations from Stevenson et al. We are able to closely recover the cloud-free results, even though `PICASO` utilizes a coarse spatial grid. We find that cloudy phase curves provide much better agreement with the WFC3 and Spitzer nightside data, while still closely matching the dayside emission. This work provides the community with a convenient, user-friendly tool to interpret phase-resolved observations of exoplanet atmospheres using 3D models.

*Unified Astronomy Thesaurus concepts:* Exoplanet astronomy (486); Extrasolar gaseous planets (2172); Hot Jupiters (753); Exoplanet atmospheres (487); Planetary atmospheres (1244); Atmospheric clouds (2180)

## 1. Introduction

Since the 1990s, more than 4000 exoplanets have been confirmed by ground- and space-based telescopes, with follow-up observations yielding insights into the complex diversity of chemical compositions, circulation patterns, and vertical structures of planetary atmospheres. Interpretation of these complex data sets requires the development of increasingly detailed and computationally efficient theoretical models, including the use of three-dimensional (3D) general circulation models (GCMs), which are able to reproduce global-scale dynamical features such as day–night temperature contrasts (especially for tidally locked planets), and equatorial super-rotation (see Showman et al. 2010, Heng & Showman 2015, Showman et al. 2020, and Zhang 2020, for reviews on the atmospheric dynamics of hot giant planets). Features such as an eastward-shifted hotspot due to strong equatorial winds were predicted in the first hot Jupiter GCMs from Showman & Guillot (2002), and were later confirmed by observations from Knutson et al. (2007). The 3D models that followed, in general, are in good agreement with these general attributes of atmospheric dynamics (e.g., Showman et al. 2009, 2015; Lewis et al. 2010, 2013; Rauscher & Menou 2010; Heng et al. 2011; Perna et al. 2012; Dobbs-Dixon & Agol 2013; Parmentier et al. 2013, 2016; Mayne et al. 2014; Kataria et al. 2015, 2016; Mendonca et al. 2016; Dobbs-Dixon & Cowan 2017; Zhang & Showman 2017; Mayne et al. 2019; Ge et al. 2020).

The use of 3D GCMs is especially valuable in interpreting phase-resolved observations (or phase curves) of exoplanet atmospheres. When conducted at infrared (IR) wavelengths, they probe an exoplanet's longitudinal thermal structure, which is controlled by the interplay of chemistry, radiation, and dynamics. Features such as the phase curve amplitude, phase offset, and wavelength dependence, yield insights into the day–night temperature contrast, hotspot shift, and the thermal and chemical structure of the atmosphere as a function of depth (Parmentier et al. 2018; Showman et al. 2020; Zhang 2020). These comparisons have been conducted for several planets using different techniques and approximations (e.g., Dobbs-Dixon & Agol 2013; Lewis et al. 2014; Zellem et al. 2014; Kataria et al. 2015, 2016; Amundsen et al. 2016; Wong et al. 2016; Lewis et al. 2017; Parmentier & Crossfield 2017; Stevenson et al. 2017; Mendonca et al. 2018a; Parmentier et al. 2018; Arcangeli et al. 2019; Flowers et al. 2019; Steinrueck et al. 2019; Parmentier et al. 2020; Venot et al. 2020; Harada et al. 2021; Roman et al. 2021). These and other theoretical modeling efforts have yielded unprecedented results and allowed for detailed constraints on the 3D processes and properties of hot Jupiters, e.g., temperature structure, the existence of magnetic drag, atmospheric metallicity, etc. (Zhang 2020).

Clouds and hazes are expected to significantly affect the chemical and vertical structures of exoplanet atmospheres, as well as their reflected, transmitted, and emitted radiation (e.g., Deming et al. 2013; Marley et al. 2013; Morley et al. 2013;

Sing et al. 2013, 2015; Stevenson et al. 2014, 2017; Mai & Line 2019; Zellem et al. 2019; Parmentier et al. 2020; Venot et al. 2020; Barstow 2021; Charnay et al. 2021). (See Marley et al. 2013, Zhang 2020, and Showman et al. 2020 for a review.) This, in turn, is expected to have a large effect on the shape and amplitude of thermal phase curves. For tidally locked planets such as hot Jupiters, one-dimensional (1D) cloud models (Ackerman & Marley 2001; Gao et al. 2018, 2020; Powell et al. 2018, 2019; Venot et al. 2020; Fraine et al. 2021) show that different gases could condense inhomogeneously on their daysides and nightsides of hot Jupiters, which can lead to complex, asymmetric 3D cloud distributions (Parmentier et al. 2016). This highlights the need for models that can flexibly investigate the properties of clouds in phase-resolved measurements.

There are several different methodologies for computing the phase curves of exoplanets. One class of codes rely on user-defined brightness temperature maps. For example, Louden & Kreidberg (2018) developed the open-source code SPIDER-MAN,[5] which relies on an arbitrary two-dimensional brightness temperature map. The integration scheme used is based on a radial coordinate system, where the area of the visible hemisphere is segmented into small regions that allow for easy calculation of the portions occulted by the star. Additionally, Luger et al. (2019) developed starry,[6] an open-source code to analytically compute the thermal phase curves of celestial bodies in occultation. Similar to SPIDER-MAN, this code requires the user to input an intensity map, and this map needs to be able to be expressed as a sum of spherical harmonics. Because these codes do not include a scheme for radiative transfer, this method prohibits the computation of spectroscopic phase curves—accounting for molecular absorption and atmospheric scattering, which are critical for interpreting Hubble Space Telescope (HST) and James Webb Space Telescope (JWST) observations.

Another way to generate phase curves is directly from the GCM radiative transfer calculation itself. One hindrance to this method is that GCMs require low-resolution grids that span a complete spectral energy distribution (SED) spectral range (e.g., 0.3–300 $\mu$m). For comparison with photometric phase curves, these methods are often sufficient (e.g., Kataria et al. 2013, 2015; Wong et al. 2015, 2016; Zhang et al. 2017). Examples of GCMs whose output has been compared to photometric phase curves include the SPARC/MITgcm, (Showman et al. 2009), the 3D radiative-hydrodynamical model from Dobbs-Dixon & Agol (2013), the GCMs used in Roman & Rauscher (2019), Roman et al. (2021), and Mayne et al. (2014), the 3D Monte Carlo Radiative Transfer (MCRT) model from Lee et al. (2019), and the GCM exo-FMS used in Hammond & Pierrehumbert (2017), Hammond & Pierrehumbert (2018), and Lee et al. (2021a).

The last class of tools alleviates the limitation of the GCM itself by relying on what is commonly referred to as "post-processing." In this schema, the GCM output is fed to an external (or internal) radiative transfer model that contains a higher-resolution cross-section grid, or the ability to compute spectra "line-by-line." In addition to enabling higher-resolution spectra, post-processing can be necessary for GCM models that have simplified radiative transfer schemes (e.g., gray or semi-gray; e.g., Roman & Rauscher 2017; Mendonca et al. 2018a; Roman & Rauscher 2019). Lastly, post-processing allows users to approximate the effect of clouds if they are not already included within the GCM calculation, or if the cloud calculation is simplified to allow for increased computational efficiency. Simplifications could include neglecting wavelength-dependent scattering (e.g., Parmentier et al. 2016; Roman & Rauscher 2019; Harada et al. 2021; Roman et al. 2021), the omission of multiple condensate species (e.g., Parmentier et al. 2020), and the exclusion of more detailed cloud microphsyics (e.g., Lines et al. 2019; Roman & Rauscher 2019; Harada et al. 2021). For these reasons, post-processing is a highly utilized methodology for computing spectra from GCMs.

There are various codes that can be used for post-processing to different extents. For example, the Planetary Spectrum Generator (PSG[7]; Villanueva et al. 2018) contains a 3D orbital calculator for computing transmission and emission spectroscopy. For direct imaging and secondary eclipse models, the code performs the radiative transfer across the visible hemisphere with adequate weights. They employ either a disk-sampling methodology that divides the observable hemisphere into sections of similar incidence/emission angles, in the form of rings centered at the subobserver point or a standard Gauss–Chebyshev disk-integration method (Saxena et al. 2021). PSG has been used for post-processing transit spectroscopy from GCM output (e.g., May et al. 2021).

In addition, the open-source radiative transfer code HELIOS[8] (Malik et al. 2017) has been used to produce phase curves from the THOR[9] GCM output (e.g., Mendonca et al. 2016). In another example, Changeat et al. (2021) updated TauREx-3 (Al-Refaie et al. 2020, 2021; Changeat & Al-Refaie 2020), a flexible exoplanet retrieval forward model, to allow for computation of the spectroscopic phase curves of the emission of tidally locked planets. In their model, they assume that the temperature variations of a 3D GCM map can be effectively captured by three concentric spherical circles made up of $n$ Gauss angles (typically $n = 8$). Therefore, they need only to calculate $3 \times 8$ individual radiative transfer equations. While this model accounts for high day–night temperature contrasts and the hotspot, it might not reproduce detailed atmospheric irregularities (e.g., Showman et al. 2009).

More recently, Lee et al. (2021b) developed an open-source GPU code, gCMCRT, which is a Monte Carlo radiative transfer forward model. The goal of gCMCRT is to automate the post-processing pipeline that computes synthetic spectra from direct global circulation model output. They post-processed results from several GCM groups including ExoRad, SPARC/MITgcm, THOR, UK Met Office UM, Exo-FMS, and the Rauscher model, and produced emission, reflection, and transmission spectra.

Lastly, Harada et al. (2021) post-processed high-resolution spectra and clouds from initial GCM results from Roman & Rauscher (2019), in addition to exploring active clouds within a GCM framework.

Given the wide usage of post-processing and the range of assumptions that it allows for, it is essential that there exist a variety of open-source methodologies capable of generating spectroscopic phase curves from 3D GCMs, conserving the inhomogeneous thermal structure of the atmosphere and with the ability to account for cloud scattering. Ultimately, this need is the motivation for developing this methodology within PICASO (Batalha et al. 2019; Batalha & Rooney 2020a).

---

[5] https://github.com/tomlouden/spiderman
[6] https://github.com/rodluger/starry
[7] https://psg.gsfc.nasa.gov/helpmodel.php
[8] https://github.com/exoclime/HELIOS
[9] https://github.com/exoclime/THOR

### 1.1. This Work

PICASO is a user-friendly, self-contained, versatile, and open-source Python-based code for modeling the reflected, transmitted, and thermal light of exoplanets in 1D and 3D. We expand this tool by adding a pipeline to map 3D GCM output onto a radiative transfer scheme that enables the computation of thermal phase curves, with the capability of including scattering by a diversity of cloud species. While PICASO has been previously used for 1D (Fraine et al. 2021) and 3D reflected light (Adams et al. 2022), and thermal emission spectroscopy in 1D (Tang et al. 2021), the work presented here represents the first test of the 3D thermal component of the code. Here we focus on thermal phase curves, but our methodology can also be easily applied to phase-resolved observations in reflected light. Because PICASO is coupled to Virga (Batalha 2020), an open-source cloud code based on the Ackerman & Marley (2001) model, we are also able to produce cloud-free and cloudy phase curves with post-processed clouds.

We benchmark our new phase curve code methodology within PICASO against spectroscopic observations and 3D models of WASP-43b, a 2 $M_J$, 1 $R_J$ exoplanet in a 19.5 hour orbit around its host star (Gillon et al. 2012). Stevenson et al. (2014) performed phase-resolved emission spectroscopy of WASP-43b with the HST Wide-Field Camera 3 (WFC3), which spans 1.1–1.7 $\mu$m. Interpretation of that data via 3D GCMs (Kataria et al. 2015, hereafter K15), found that cloud-free models closely matched the dayside observations but overpredicted the flux on the nightside. This disagreement in the nightside flux could be explained by the existence of clouds, which decrease the emergent flux and restrict observations to higher, cooler altitudes. We produce cloud-free phase curves to benchmark against the cloud-free phase curves in K15, and produce cloudy phase curves to compare to the HST/WFC3 observations from Stevenson et al. (2014). This work lays out a user-friendly framework that can be used to predict and interpret future thermal emission observations of transiting and directly imaged exoplanets, such as those planned for the JWST.

In Section 2 we describe the models used and the new update to PICASO. In Section 3 we present the cloud profiles computed with Virga, as well as our synthetic cloudy and cloud-free thermal phase curves and spectra of WASP-43b. We also discuss the results, reflect on the utility of the new model, and make predictions for future JWST observations. In Section 4 we lay out our conclusions.

## 2. The Models

### 2.1. SPARC/MITgcm

The cloud-free and cloudy phase curves we produce using PICASO and Virga are derived from the 3D temperature and wind profiles presented in K15. These fields were produced using the SPARC/MITgcm, a model that couples the MITgcm (Adcroft et al. 2004) to a two-stream adaptation of the multistream radiative transfer code by Marley & McKay (1999). The MITgcm is a GCM used to simulate atmosphere

**Table 1**
Properties of the WASP-43A/b System

| Parameter | Value |
| --- | --- |
| $M_* [M_\odot]$ | 0.717 |
| $R_* [R_\odot]$ | 0.667 |
| $M_P [M_J]$ | 2.034 |
| $R_P [R_J]$ | 1.036 |
| $a$ [AU] | 0.01526 |
| $T_{\rm orb} = T_{\rm rot}$ [days] | 0.81 |
| $g$ [m s$^{-1}$] | 47.0 |

**Note.** From Gillion et al. (2012).

and ocean dynamics (Adcroft et al. 2004). It uses the finite volume discretization method to solve the primitive equations for a stably stratified atmosphere on a cubed sphere grid: the horizontal and vertical momentum, continuity, and thermodynamic energy equations. These are a simplification of the Navier–Stokes equations, since hot Jupiters have generally highly stratified atmospheres and the horizontal scale greatly exceeds the vertical scale when studying global flows.

Tidally locked planets consist of a hot dayside (facing the star) and a colder nightside (facing away from the star). This strong temperature gradient causes winds to move the atmosphere away from radiative equilibrium. The SPARC/MITgcm self-consistently calculates the heating rate from the radiative transfer (Showman et al. 2009). The latter computes the upward and downward fluxes at each atmospheric layer through all the vertical columns in the GCM. These fluxes are then used to calculate and update the wind velocities, temperatures, and heating/cooling rates, which in turn affect the radiative transfer. The opacities are divided into 11 frequency bins using the correlated-$k$ method (Goody et al. 1989; Kataria et al. 2013), which is computationally efficient and retains most of the accuracy of full line-by-line calculations. K15 assume chemical equilibrium (accounting for rainout of condensates) at any local pressure and temperature, and they use the elemental abundances in Lodders (2003) to calculate the opacities, ignoring clouds and hazes. For more details on this model, see, for example, Showman et al. (2009), Kataria et al. (2013), and Kataria et al. (2015).

In this study we derive cloudy and cloud-free phase curves and spectra from K15, who modeled WASP-43b with a C32 resolution grid (global resolution of 128 × 64 in longitude and latitude), with 40 evenly spaced log pressure layers between 0.2 mbar and 200 bar. Here we assume an atmospheric metallicity of 1× solar, but it is possible to conduct the same analysis using 5× solar metallicity as done in K15 (this is left for future work). We use the 3D GCM temperature maps from K15, as a function of pressure, as inputs for the post-processed radiative transfer computation to obtain the outgoing thermal emission. The WASP-43A/b system parameters adopted for this model are shown in Table 1.

### 2.2. PICASO

PICASO is an open-source Python-based radiative transfer model for computing the reflected, thermal, and transmitted light of exoplanets (Batalha et al. 2019). The PICASO code is publicly available on Github[10] and includes step-by-step tutorials on how to perform the different types of calculations, as well as an in-depth description of the radiative transfer equations and mathematical methods used in the model. The methodology of PICASO to compute the radiative transfer for thermal emission is partly rooted in the Fortran self-consistent radiative-convective model described by McKay et al. (1989) and Marley & McKay (1999), for brown dwarfs and solar system planets. This atmospheric modeling code was also used by other groups for exoplanet atmospheres (e.g., Fortney et al. 2005, 2006, 2008, 2010; Kataria et al. 2013, 2014, 2015, 2016; Morley et al. 2013, 2015; Parmentier et al. 2013).

---
[10] https://github.com/natashabatalha/picaso

PICASO takes the physical properties of the planet (such as its mass and radius) and the host star (mass, radius, temperature, metallicity, and stellar flux) as the basic inputs. We computed post-processed chemical equilibrium calculations at every point using the CEA code (Gordon & McBride 1994), given the pressure–temperature grid from K15. PICASO uses the opacities from the Resampled Opacity Database for PICASO v2 (Batalha et al. 2020b). The opacities that we used for the post-processed radiative transfer (outlined in Marley et al. 2021) are nearly the same as the ones used in the SPARC/MITgcm, on different wavelength resolution grids. However, these are updated opacities, versus those that were used in the original GCM calculations of K15 (outlined in Freedman et al. 2014). To ensure that the post-processed spectrum is consistent with the thermal equilibrium spectrum reached by a GCM, users should strive to use the same set of opacities.

The 3D functionality in PICASO is all based on the output format of MITgcm models, though we encourage users to submit requests for different input formats. Currently, we include a convenient function to regrid 3D temperature and eddy mixing coefficient ($K_{zz}$) maps from the original-resolution MITgcm grid onto the desired new, integrable sphere grid. $K_{zz}$ maps are relevant to determine vertical mixing, which in turn affects the cloud profiles and particle-size distributions; they are thus important for computing cloud properties with Virga (see Section 2.3).

The original thermal emission methodology (Marley & McKay 1999) relied on spherical symmetry and used eight Gauss angles to compute the disk-integrated flux. Following the methodology of Horak & Little (1965), which was utilized in the reflected-light component of PICASO (Cahoy et al. 2010; Batalha et al. 2019), we implement a Chebyshev–Gauss integration scheme to compute the disk-integrated flux. The "Chebyshev" component of this scheme effectively adds the latitude dependence that we need to pair the GCM results to our radiative transfer scheme. Therefore, the combined Chebyshev–Gauss angles provide the outgoing angles needed to compute the emergent intensity at each disk facet (creating a "disco ball"-like integration grid of the visible sphere). Our "disco ball" methodology effectively requires $n$ Gauss $\times$ $t$ Chebyshev angle radiative transfer computations.

PICASO computes the flux at every layer and angle in the grid following the two-stream source function procedure outlined in Toon et al. (1989). The flux is integrated for all layers to obtain a thermal flux at the top of the atmosphere. The emergent planetary flux is integrated over the visible hemisphere using the Chebyshev–Gauss integration method outlined in Horak (1950) and Horak & Little (1965), also used by Cahoy et al. (2010), Madhusudhan & Burrows (2012), and Webber et al. (2015).

The general expression for the ratio of the planet flux and stellar flux as seen by an observer for every visible hemisphere is adapted from Rauscher et al. (2018; see also Cowan & Agol 2008; Cahoy et al. 2010; Cowan & Agol 2010; Cowan et al. 2013) as

$$\frac{F_P(\xi)}{F_*} = \left(\frac{R_P}{R_S}\right)^2 \frac{\oint V_P(\theta, \phi) M_P(\theta, \phi, \xi)\, d\Omega}{\oint V_*(\theta, \phi) M_*(\theta, \phi)\, d\Omega}, \quad (1)$$

where $R_P$ and $R_S$ are the planet and stellar radius, $V_P$ and $V_S$ are the area weighting functions, $M_*$ is the outgoing stellar flux, and $M_P$ is the top of the atmosphere flux. Here $M_P(\theta, \phi, \xi)$ depends on the orbital phase $\xi$, but for thermal emission all calculations are done by setting the phase to 0 in PICASO. This is further discussed in Section 2.4. Even at phase 0, the viewing angles depend on longitude and latitude across the disk, so the viewing angle is varied at each longitude–latitude pair. $\theta$ and $\phi$ are latitude and longitude, and this equation represents an integration over a solid angle, $d\Omega = \cos(\theta) d\theta d\phi$. Thus, the flux is weighted by the cosine of the latitude, which lowers the contribution of high latitudes to the total flux calculation. We assume that the stellar flux is constant in time and throughout the orbit, so $\oint V_S d\Omega = \pi$. The original expression can then be rewritten as

$$\frac{F_P(\xi)}{F_*} = \left(\frac{R_P}{R_S}\right)^2 \oint V_P(\theta, \phi) \frac{M_P(\theta, \phi, \xi)}{\pi M_*}\, d\Omega. \quad (2)$$

In PICASO, $V_P(\theta, \phi)$ and $\cos(\theta)$ are accounted for by the weights for each longitude and latitude point on the Chebyshev–Gauss grid. These specific angles are the same for all points in the orbit for a constant grid resolution.

This integration is performed at every wavelength in the desired wavelength range, and yields the 1D compressed thermal flux as a function of wavelength, i.e., the thermal emission spectrum.

### 2.3. Virga

We investigate the effects of clouds on synthetic thermal phase curves and phase-resolved thermal spectra using the Virga[11] open-source cloud model developed by Batalha (2020). Virga computes the optical depth and scattering properties of the condensate clouds for hot Jupiters and other substellar objects. Clouds are described by a single scattering albedo, an asymmetry parameter, and a total extinction per layer (optical depth), all as a function of pressure and wavelength. The refractive indices and Mie scattering parameters for different condensate species are drawn directly from Batalha & Marley (2020b). Virga calculates cloud particle-size distribution and the effect of vertical mixing and sedimentation on cloud shapes and locations using the methodology described in Ackerman & Marley (2001; see, e.g., Morley et al. 2012, for an application to brown dwarfs), for each condensate independently, and ignoring any microphysical interactions between clouds. In order to compute cloud profiles, Virga takes as input temperature and $K_{zz}$ profiles, estimated by assuming $K_{zz} = \omega(z) L(z)$, where $\omega(z)$ is the globally averaged rms vertical velocity at a given pressure level, taken from the MITgcm, and the mixing length $L(z)$ is approximated as the atmospheric scale height $H(z)$ (Adams et al. 2022). Limitations of this approach are discussed in Parmentier et al. (2013), Zhang & Showman (2018), Komacek et al. (2019), and Menou (2019). `Virga` also intakes a value of sedimentation efficiency ($f_{sed}$) as an adaptable parameter. Ackerman & Marley (2001) define $f_{sed}$ as the efficiency of the atmosphere to deposit cloud particles toward higher pressures, or the ratio of the mass-weighted droplet sedimentation velocity to the convective velocity scale. Their model balances upward transport of cloud condensates driven by vertical turbulent mixing with the deposition of these condensates through sedimentation. The balance equation can be written as

$$-K_{zz}\frac{\partial q_t}{\partial z} - f_{sed} w_* q_c = 0, \quad (3)$$

where $K_{zz}$ is the vertical eddy diffusion coefficient, $w^*$ is the convective velocity scale, $q_t$ is the total mixing ratio of condensate and vapor, and $q_c$ is the mixing ratio of condensates only. This equation is solved for each condensate species independently. If $K_{zz}$ is large, there is strong vertical mixing, which causes a large updraft velocity. The lower bound of $K_{zz}$ is set to be $10^5$ cm$^2$ s$^{-1}$ automatically by Virga, in order to account for the residual turbulence from breaking buoyancy waves in the atmosphere's radiative regions (Ackerman & Marley 2001; Morley et al. 2014).

---

[11] https://github.com/natashabatalha/virga

Sedimentation efficiency values are thought to be less than 1 for warm Neptunes/hot Jupiters (Ackerman & Marley 2001; Morley et al. 2013, 2015). Values of nearly 10 have been suggested for some brown dwarfs (Saumon & Marley 2008). These values are not physically motivated for both brown dwarfs and exoplanets, but instead set by observations (Gao et al. 2018). We chose an $f_{sed}$ range from 0.01 to 0.3 based on previous efforts to reproduce observations with the Ackerman & Marley (2001) cloud model (e.g., Webber et al. 2015; Christie et al. 2021). Higher values of sedimentation efficiency represent clouds that are thinner and more depleted with larger particles, while smaller values correspond to clouds that extend higher and deeper in the atmosphere with smaller particles. The cloud condensate concentration generally relates to $f_{sed}$ as

$$C(z) = C_{base} \exp\left(\frac{-f_{sed} z}{H}\right), \quad (4)$$

where $C(z)$ is the condensate concentration as a function of depth $z$, $C_{base}$ is the concentration at the cloud base, and $H$ is the local scale height (since we run `Virga` at every latitude/longitude point).

`Virga` currently supports the following cloud condensate species: $Al_2O_3$, $CH_4$, Fe, KCl, $H_2O$, $NH_3$, Cr, $Mg_2SiO_4$, $MgSiO_3$, MnS, $Na_2S$, $TiO_2$, and ZnS (see Batalha & Marley 2020b). Cr, Fe, $Mg_2SiO_4$, $MgSiO_3$, MnS, and $Na_2S$ have been hypothesized to dominate the cloud condensate mass in exoplanet atmospheres (Lodders 1999; Visscher et al. 2006, 2010; Gao et al. 2020; Venot et al. 2020). We assume solar abundance ($1\times$ solar) of the limiting elements. For the purposes of this work, we select the cloud species recommended by `Virga` for WASP-43b's temperature and $K_{zz}$ maps at every point in the orbit. `Virga` recommends species by comparing their condensation curves to the temperature profile at every longitude–latitude point in the atmosphere. The assumption in this step is that a species will condense when its partial pressure is greater than the vapor pressure, and thus `Virga` needs to compute the available quantity of cloud condensates. Other factors come into play to determine whether a species will condense, such as vertical mixing and the chemical composition of the atmosphere.

### 2.4. Computing Thermal Phase Curves with `PICASO`

`PICASO` already has the methodology built in to compute 1D thermal emission derived from 3D models (see Section 2.4), which can be compared with other models and observational data. However, until now, the code could only account for different phase geometry in reflected-light calculations. Inputting different phase angles into the code[12] would only compute the flux for the visible crescent of the planet that was visible in reflected light at every orbital phase. But planets emit thermal emission at all latitudes and longitudes, regardless of incoming angle. The flux variation that is seen is instead a function of the cooler regions of the planet observed throughout the planet's orbit. We added a new pipeline to `PICASO` that allows users to compute the phase-resolved thermal emission of 3D planetary atmospheres, with and without clouds.

To perform thermal phase curve calculations, we need to input the 3D pressure–temperature and pressure–$K_{zz}$ maps (the latter for `Virga` calculations) that will allow us to compute the radiative transfer for visible longitudes at each orbital angle. To do this, we simply rotate the 3D grid by the orbital phase angle, at every point in the orbit, to center the temperature map at the subobserver (Earth-facing) longitude. The latitude coordinates remain unchanged. WASP-43b is a good candidate to test our phase curve code because it is assumed to be tidally locked in a nearly circular orbit. This makes the rotation period of the planet equal to the orbital period, which greatly simplifies our methodology to determine the Earth-facing longitude at each point in the orbit. The wealth of data that exists for this planet also allows for close comparisons between the model and observations.

Transferring the full GCM grid to the integrable angles makes up the first step in setting up the 3D thermal emission calculation. `PICASO` performs an area-weighted average of the pressure, temperature, and $K_{zz}$ maps by binning the GCM longitude and latitude points found between adjacent points of the Gauss–Chebyshev grid (selected by a nearest-neighbor query). We regrid the temperature and $K_{zz}$ maps from the K15 SPARC/MITgcm output at a $128 \times 64$ resolution, down to a $10 \times 10$ resolution. Ultimately, this corresponds to 100 individual radiative transfer calculations. For its phase-dependent reflected-light component, `PICASO` chooses 10 Gauss and 10 Chebyshev angles as a default (e.g., Adams et al. 2022). In order to determine the default number of angles we ran simulations for angular resolutions of $10 \times 10$, $15 \times 15$, and $20 \times 20$. Ultimately, we chose a $10 \times 10$ resolution for our analysis and results because it is able to reproduce the longitudinal thermal structure of the atmosphere computed by K15 with the $128 \times 64$ grid, as seen in Figure 1. It is worth noting that, with this grid, we lose information about the thermal flux at the planetary limbs, as well as the poles. From our sensitivity calculations, there is negligible information in these angles for the case explored in this work. An increase in spatial resolution (e.g., 20 × 20) leads to much higher computation time (which increases even more when adding clouds) and negligible differences in the resulting thermal emission, considering the large error inherent in the data (see Appendix). However, a higher-resolution grid would be needed for an investigation where the planet limbs were expected to contribute significantly to the total flux. Figure 1 illustrates the 10 × 10 grid, for which the Gauss–Chebyshev angles correspond to longitudes between −76°.9 and 76°.9 and latitudes between −73°.6 and 73°.6 from the subobserver point (directly facing Earth).

---

[12] Via the https://natashabatalha.github.io/picaso/notebooks/8_Spherical Integration.html##Geometry-with-Reflectivity-and-Non-Zero-Phase phase_angle routine.

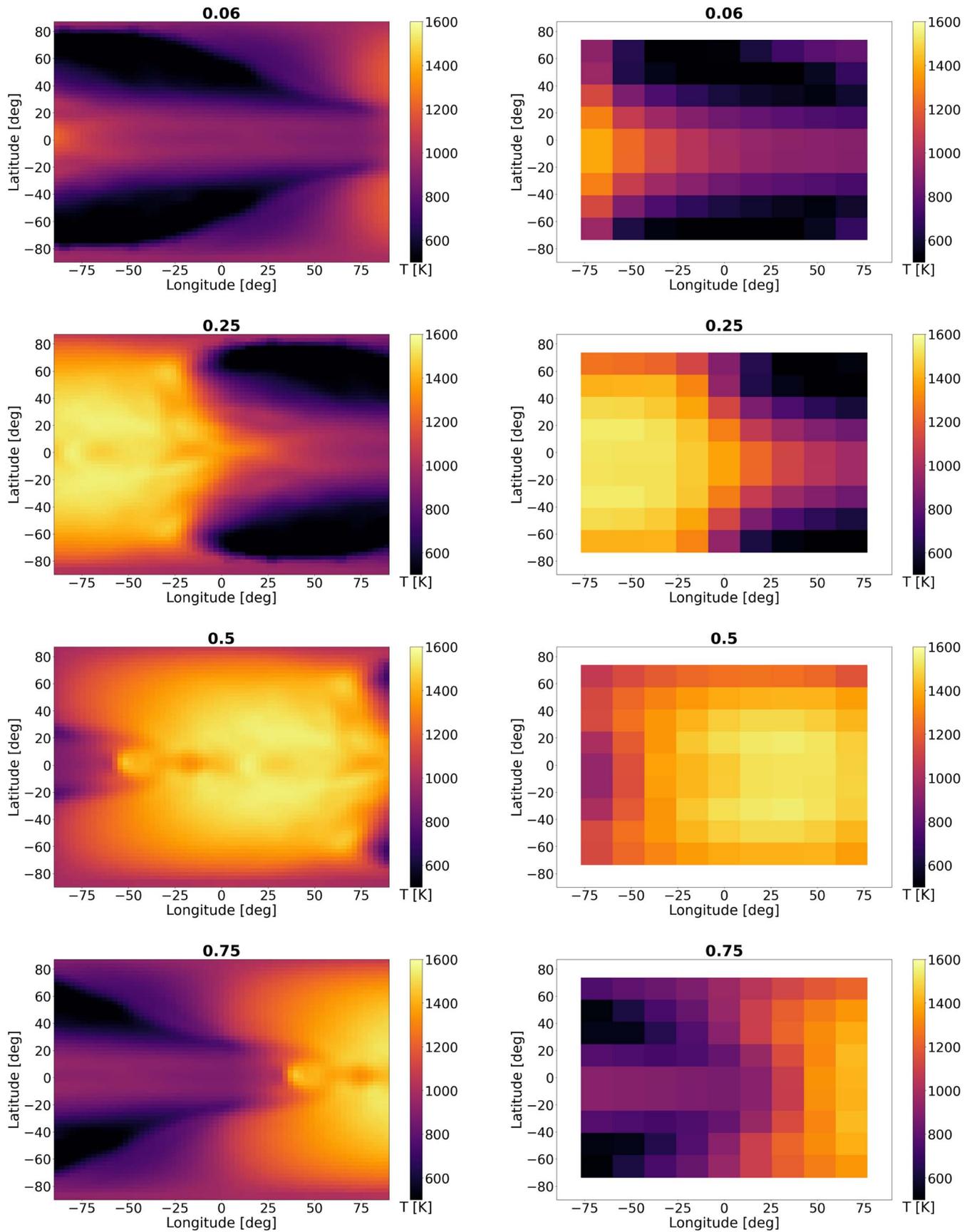

**Figure 1.** Temperature maps of the visible hemisphere at various orbital phases of WASP-43b, plotted at a pressure of 10 mbar. Left column shows the MITgcm native 128 × 64 resolution, while the right column plots the temperature on a 10 × 10 Gauss–Chebyshev grid. Note the white areas in the maps on the right, which are due to the coarser-resolution grid; the 10 × 10 Chebyshev–Gauss grid only computes the flux for longitudes from −76°.9 to 76°.9 and for latitudes from −73°.6 to 73°.6.

Once the geometry is taken care of, PICASO computes the outgoing thermal emission flux for every visible hemisphere in the orbit. The methodology to generate synthetic spectra and phase curves in PICASO is similar to the procedure used by K15 (Fortney et al. 2005, 2006, 2008, which are both derivatives of Marley & McKay 1999). Scattering due to clouds is included into the radiative transfer calculation by inputting the cloud profiles from Virga (single scattering albedo, asymmetry parameter, and total extinction).

Computing the post-processed radiative transfer is an efficient way of making predictions and interpretations of future observations by the JWST. While the coupling of PICASO and Virga is not energetically self-consistent (temperatures are not updated by the presence of clouds), it provides a useful tool to test a rich parameter space of cloud properties without the need to rerun the full GCM, and see the effects of clouds on the resulting phase curves and spectra.

To compute phase-resolved emission, we choose a phase resolution of 15 points around the orbit, from 21°.6 to 338°.4 (as chosen in Stevenson et al. 2014). Transit occurs at 0° (or 360°), corresponding to the normalized phase 0. Secondary eclipse occurs at 180° after transit, corresponding to phase = 0.5 in the plots. We find that this orbital angle resolution is enough to recover the shape of the phase curve and is relatively fast to run for a 10 × 10 grid.

## 3. Results and Discussion

### 3.1. Clouds on WASP-43b

Using the 3D pressure–temperature and pressure–$K_{zz}$ grids from SPARC/MITgcm as input, we first run Virga across the 10 × 10 grid of the visible hemisphere, at every point in the orbit of WASP-43b, and for multiple values of $f_{sed}$. For every point in this 3D grid, Virga computes the size of the cloud particles and the optical properties: single scattering albedo, column optical depth (total and for individual species), and asymmetry parameter. Figure 2 shows the hemisphere-averaged column optical depth for $f_{sed} = 0.3$ at orbital phases shown in Stevenson et al. (2014), corresponding to the nightside (phase = 0.06, 0.12, 0.19, 0.81, 0.88, 0.94), limbs (phase = 0.25, 0.75) and dayside (phase = 0.31–0.69). We choose $f_{sed} = 0.3$ in this analysis because it provides a close fit to observations (see Section 3.2). These optical-depth calculations assume geometrically spherical particles. The geometric column optical depth is inversely proportional to the magnitude of $f_{sed}$, so that increased precipitation reduces opacity by decreasing the mixing ratio of condensates, $q_c$, and increasing the effective droplet radius (Ackerman & Marley 2001). The turnover point of the solid lines, corresponding to the column optical depth, represents the base of the cloud (Morley et al. 2014). The exact vertical extent of the cloud depends on the interaction between vertical mixing and sedimentation (Parmentier et al. 2020).

Figure 2 shows the column optical depth for each gas integrated from the top of the atmosphere. The per-gas optical depths in Virga are computed assuming a geometric scattering approximation. This conservative optical-depth computation assumes the particles are large and have a constant extinction coefficient. They are only used for visual purposes to illustrate the breakdown of the condensate contributions. An overall higher optical depth can be seen for all species near the nightside of the planet. On the colder nightside, more species condense to become optically thick. On the hotter dayside, we see that clouds tend to form deeper in the atmosphere. Some species, such as $MgSiO_3$ (enstatite), $Mg_2SiO4$ (fosterite), MnS, $Na_2S$, ZnS, and Cr, are present throughout all planet longitudes. Venot et al. (2020) showed that both $MgSiO_3$ and MnS are able to sufficiently dim the nightside emission spectrum shortward of 5 $\mu$m, and concluded that MnS clouds would only form on the nightside. $Na_2S$ exhibits an optically thin cloud. Fe is present in almost every orbital phase at pressures above 3 bars, decreasing as the nightside becomes visible and the temperatures cool. $Al_2O_3$ clouds are only present at deep layers of the dayside, due to their high condensation temperature. Cr clouds were found to provide enough opacity to increase the phase curve amplitude in the HST/WFC3 bandpass, and are thought to be significantly opaque at Spitzer wavelengths (Parmentier et al. 2020). Fosterite and enstatite both contain the same limiting element, Mg, and have very similar condensation curves. Venot et al. (2020) and Roman et al. (2021) chose to model only fosterite for WASP-43b, arguing that fosterite would condense first due to its higher condensation temperature, depleting Mg and $SiO_2$ from the atmosphere and causing the enstatite cloud that forms above to have much lower mass. In this work, we chose to split the solar composition amount of Mg atoms between $MgSiO_3$ and $Mg_2SiO_4$. Further tests could be done to determine the individual effect of each species on the thermal emission. These results are largely the same as those obtained by Helling et al. (2020) with SPARC/MITgcm and a more complex microphysical kinetic model applied to WASP-43b.

In addition to analyzing the profile of each cloud species, it is also instructive to determine their cumulative effects. Figure 3 shows the cumulative optical depth per layer integrated over all condensate species as a function of longitude and pressure, at a latitude near the equator, for $f_{sed} = 0.3$, and for a wavelength of 1.4 $\mu$m (i.e., within the WFC3 bandpass). These are the optical depths used for the radiative transfer calculations. They are computed using Mie theory and integrated over the log normal radius distribution. The substellar longitude (facing directly toward the star) at each phase is represented by the red dashed lines, and the orange dashed lines correspond to the antistellar point (facing directly away from the star). Optically thick clouds can be seen for most longitudes near the nightside, compared to the thinner clouds near the dayside. The upper cloud layer extends down to around 1 bar, and appears to be more opaque at longitudes west of the substellar point. This is consistent with model results from Parmentier et al. (2016), which showed that the coolest regions of the dayside tend to be toward the western edge due to the eastward-flowing air mass still being cold from the nightside. Thus clouds tend to form on the nightside, persist on the western edge of the dayside, and evaporate on the dayside (e.g., Mendonca et al. 2018a; Lines et al. 2018; Powell et al. 2018; Roman & Rauscher 2019). This tendency might vary for individual planets, and it would depend on the sedimentation efficiency of the clouds, as well as their microphysical properties.

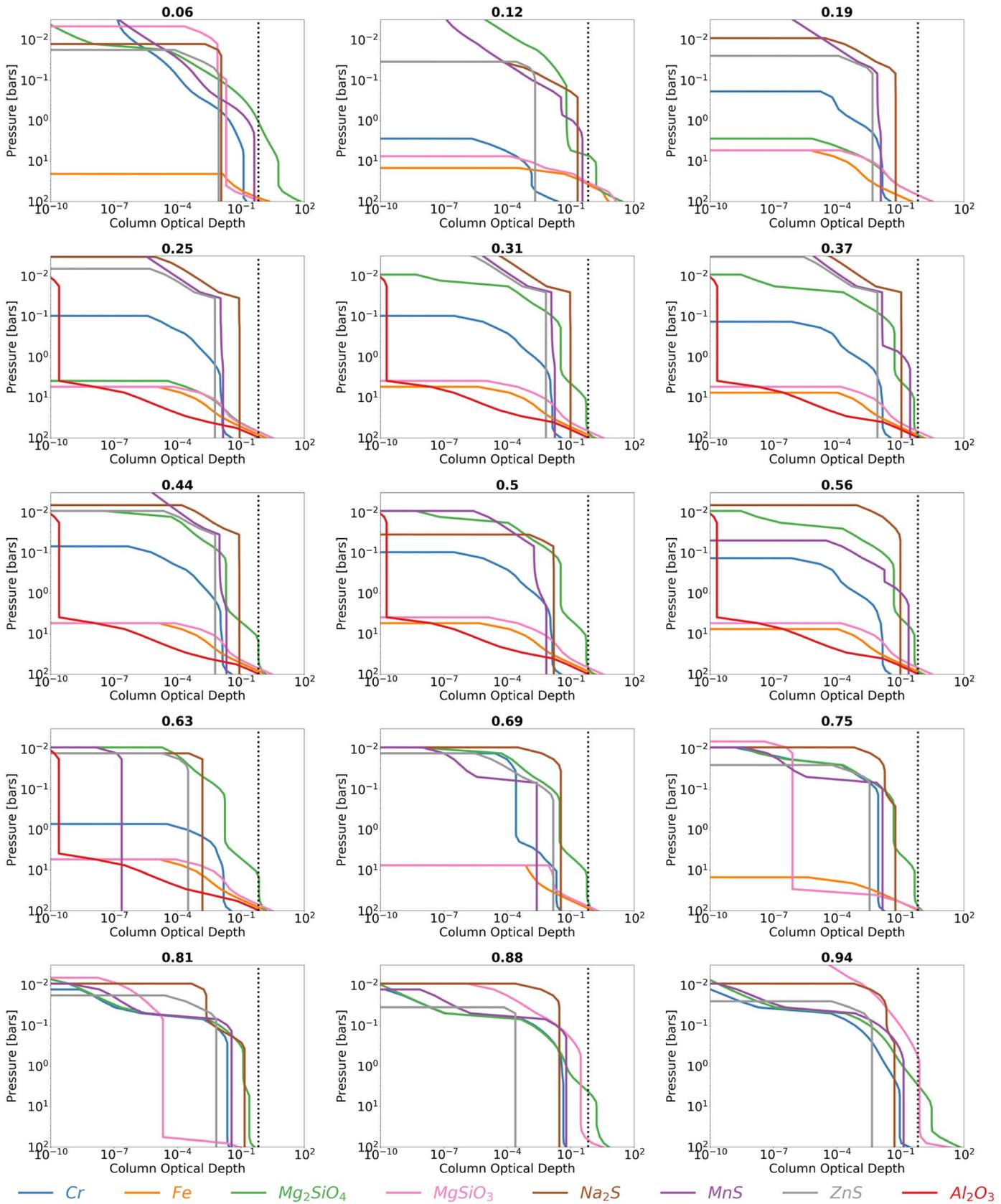

**Figure 2.** Hemisphere averages of the column optical depth for all the cloud species recommended by Virga, with $f_{sed} = 0.3$. A black dotted line is placed at the optical depth of 2/3. Each figure corresponds to a visible hemisphere along the orbit. The orbital phase angle is normalized, with the nightside visible at phases = 0.06, 0.12, 0.19, 0.81, 0.88, 0.94, limbs at phases = 0.25, 0.75, and dayside at phases = 0.31–0.69. Transit occurs at phase 0 and secondary eclipse at phase 0.5.

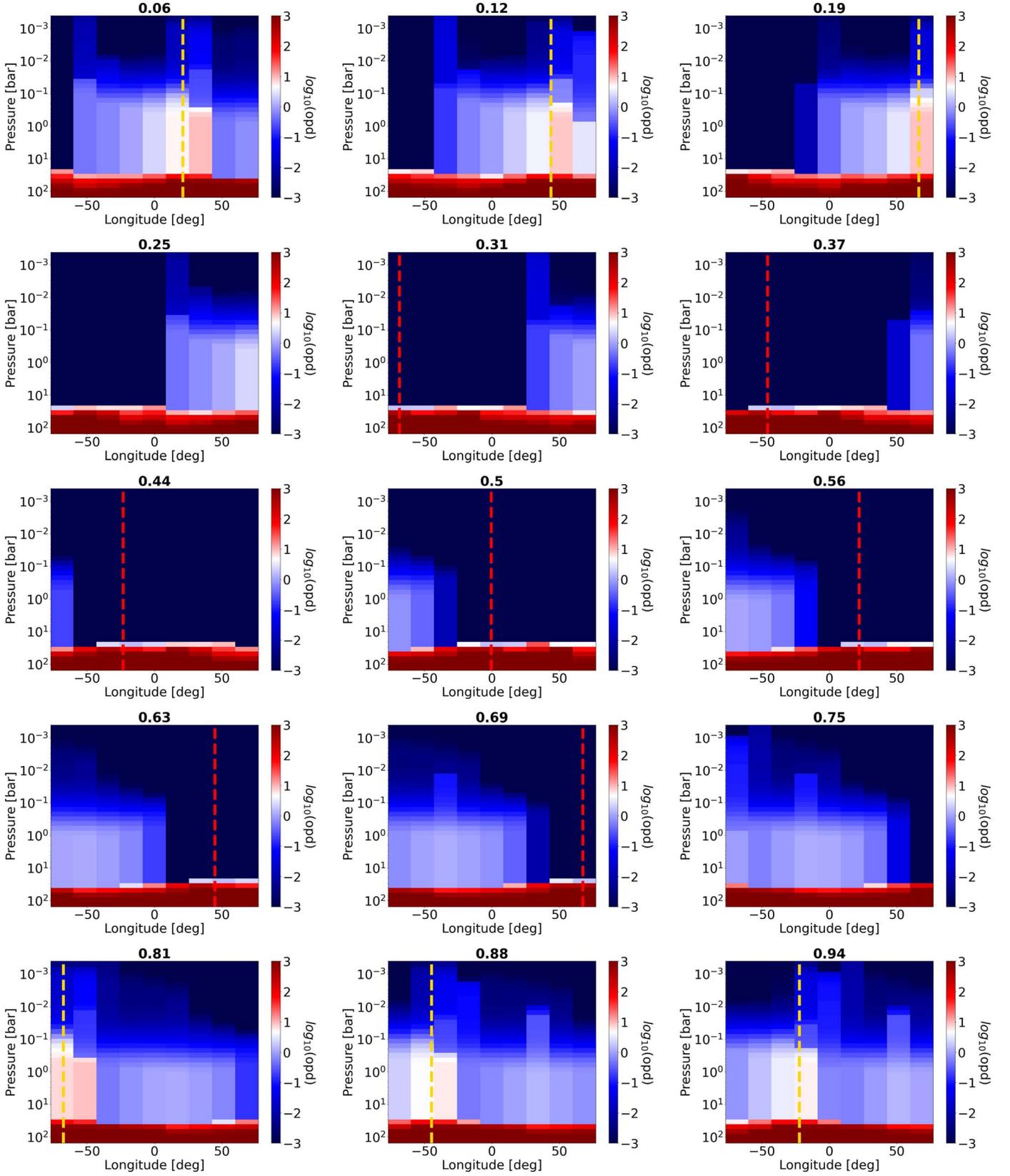

**Figure 3.** Cloud cumulative optical depth per layer integrated for all condensate species as a function of longitude and pressure, at a latitude of $-8°.18$, a wavelength of 1.4 $\mu$m, and $f_{sed} = 0.3$. The color scale is centered at an optical depth of 2/3. The orbital phase angle is normalized, with the nightside visible at phases = 0.06, 0.12, 0.19, 0.81, 0.88, 0.94, limbs at phases = 0.25, 0.75, and dayside at phases = 0.31–0.69. Transit occurs at phase 0 and secondary eclipse at phase 0.5. Each panel is centered at the corresponding phase angle, such that the 0 longitude is the Earth-facing longitude. The substellar longitude at each phase is represented by the red dashed lines, visible between phases 0.25 and 0.75, and the gold dashed lines correspond to the antistellar point, visible between phases after 0.75 and before 0.25.

We find a second opaque cloud layer at depths near 30 bars that extends down to the bottom of the grid. Deep clouds could exist on the dayside, depending on whether silicates are able to form clouds at depth, which in turn is subject to the microphysics, atmospheric dynamics, and temperature of the planet's interior layers (Venot et al. 2020). This deep cloud likely has a much smaller effect on the thermal emission observed, since the pressure levels probed by HST/WFC3 range from around 10–100 mb. A deep cloud could act as a "vertical cold trap," depleting the atmosphere of the cloud making atoms (e.g., Spiegel et al. 2009; Parmentier et al. 2013; Beatty et al. 2017; Powell et al. 2018).

### 3.2. Phase-resolved Thermal Emission

As a first test of the new thermal phase curve code with PICASO, we compute the phase-resolved thermal emission of WASP-43b derived from K15 model grids with and without clouds using Virga, and compare our model to previous HST/WFC3 observations using the G141 grism (1.1–1.7 $\mu$m; Stevenson et al. 2014). While PICASO provides a grid of stellar models, we used the same WASP-43A stellar model in K15 for closer comparison. For our cloudy models, we choose four different values of sedimentation efficiency: 0.01, 0.03, 0.1, and 0.3. The clouds included correspond to the condensation species recommended by Virga discussed in the previous section.

#### 3.2.1. White-light Phase Curves

Figure 4 shows the simulated solar-metallicity white-light phase curve with PICASO, integrated over the WFC3 bandpass. We compare this phase curve to the cloud-free models from K15. While K15 presented only cloud-free models, here we can use PICASO and Virga to include the post-processed effects of clouds.

As expected, our cloud-free model gets remarkably close to the phase curve from K15, because of our sensitivity tests which chose the spatial resolution. The lower resolution used in PICASO results in much lower computation time. Similar to the K15 results, the maximum amplitude of our cloud-free white-light phase curve is consistent with, though slightly lower than, the peak IR flux seen in the observations. We attribute the lower flux to a combination of updated opacities which are outlined in detail in Marley et al. (2021) and new chemistry calculations used in PICASO. Our PICASO cloud-free phase curve also exhibits a peak in IR flux that occurs 20°–30° after secondary eclipse (phase = 0.5), consistent with the K15 temperature maps for WASP-43b.

We find that our cloud-free PICASO phase curve overestimates the nightside flux, as well as the flux ratios for orbital phases leading from transit to secondary eclipse. Our cloudy models (blue curves) provide a much better fit for observations of the nightside, and the models with $f_{sed}$ values of 0.1 and 0.3 are better fits for phases after transit and before secondary eclipse. This is consistent with predictions for the presence of optically thick clouds in the colder regions of WASP-43b (e.g.,

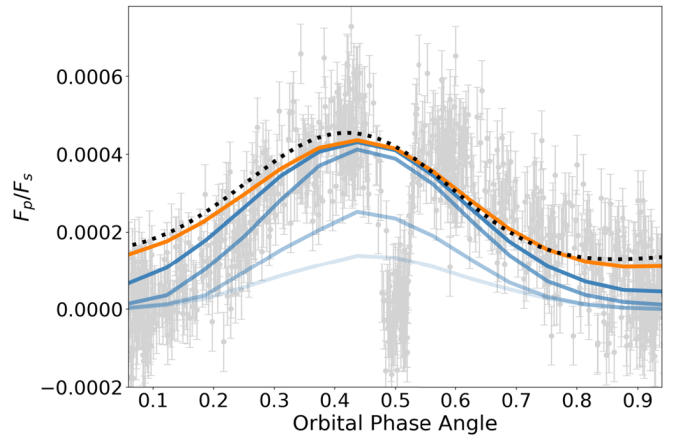

**Figure 4.** Band-integrated phase curves (white light) of WASP-43b, showing the planet-to-star flux ratio as a function of orbital phase for a 1× solar-metallicity model. The flux ratio is multiplied by the planet-to-star radius ratio $\left(\frac{R_P}{R_S}\right)^2$. The cloud-free model computed with PICASO is represented as an orange line, and the cloudy models correspond to the blue lines. The increasing opacity of the line scales with increasing values of the sedimentation efficiency $f_{sed}$ (0.01, 0.03, 0.1, and 0.3). Clouds are formed from condensation species recommended by Virga, and change for each visible hemisphere. The black dotted line corresponds to the 1× solar cloud-free model from K15, and the observational data was obtained from WFC3 (Stevenson et al. 2014). The nightside is visible at phases = 0.06, 0.12, 0.19, 0.81, 0.88, 0.94, the limbs at phases = 0.25, 0.75, and the dayside at phases = 0.31–0.69.

Stevenson et al. 2014; Kataria et al. 2015; Stevenson et al. 2017; Venot et al. 2020). Overall, cloudy models with these sedimentation efficiency values match the observations better than the cloud-free phase curve. Phase curves with low $f_{sed}$ values (0.01 and 0.03) represent very optically thick clouds with small particle sizes that are more vertically extended. As a result, these clouds greatly inhibit the outgoing thermal flux at nightside longitudes. Clouds shift the nightside photosphere to lower pressures, altering the brightness temperatures that can be probed during observations (Dobbs-Dixon & Cowan 2017; Parmentier et al. 2020; Roman et al. 2021). Near the dayside, the addition of clouds (with $f_{sed}$ values of 0.1 and 0.3) does not affect the outgoing thermal emission as much as on the nightside, indicating that thicker, more opaque clouds are forming at higher altitudes on the nightside.

Including clouds results in an increase of the phase curve amplitude and a decrease in the offset of the phase curve peak (Figure 4). This is consistent with results from Parmentier et al. (2020) and Roman et al. (2021). In Figure 5 we show band-integrated planet-to-star flux ratio maps for the cloud-free and cloudy dayside models (phase = 0.5). Even though the phase curve maximum changes between models, the dayside flux distribution is not significantly altered, except for the case with a low $f_{sed}$ value of 0.03, where clouds strongly diminish the dayside emission. It is possible that the lack of emission from the cloudy nightside at all other phases is shifting the phase curve maximum (Parmentier et al. 2020). Parmentier et al. (2020) suggested that the reason for cloudy thermal phase curve offset and large amplitude is to be found in the effects of clouds on the observed spectrum rather than the effect of their radiative feedback on the climate. This reinforces the utility of using PICASO and Virga (as a post-processing model that couples clouds with the radiative transfer) versus a fully self-consistent dynamical model with clouds, to more quickly investigate the effects of clouds on light curves.

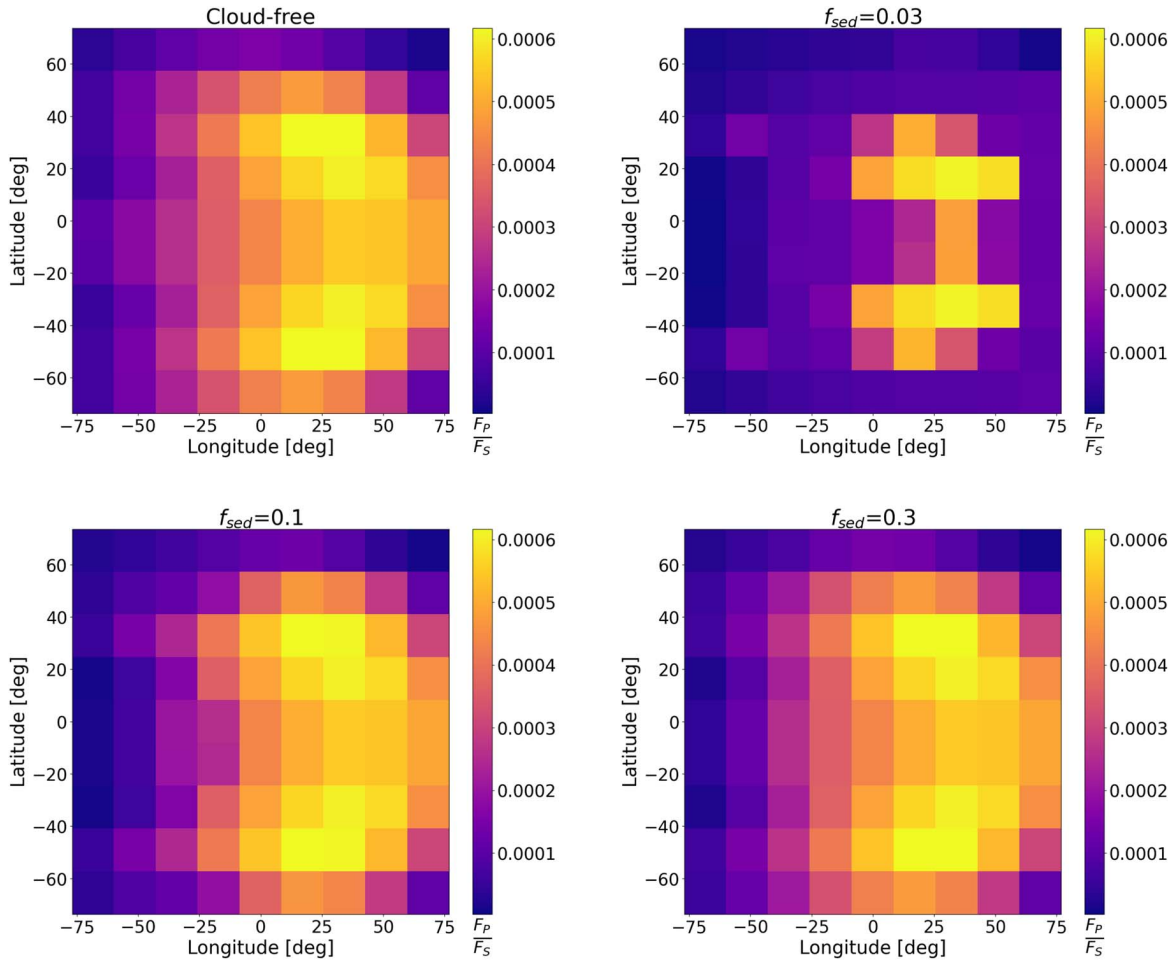

**Figure 5.** Broadband (convolved over WFC3 G141 filter) dayside (phase 0.5) relative flux maps of WASP-43b, multiplied by the planet-to-star radius ratio $\left(\frac{R_P}{R_S}\right)^2$, for the cloud-free and cloudy models with different values of $f_{\rm sed}$.

### 3.2.2. Spectroscopic Phase Curves and Phase-resolved Spectra

In addition to a white-light (band-integrated) phase curve, we are also able to produce cloudy and cloud-free spectroscopic phase curves and phase-resolved spectra with PICASO. Resolving the phase offset as a function of wavelength in a spectroscopic phase curve allows for the recovery of information about the heat transport as a function of atmospheric depth (Louden & Kreidberg 2018). Figure 6 shows spectroscopic phase curves of WASP-43b with and without clouds, using the same wavelength bins presented in Stevenson et al. (2014) and K15. Our cloud-free model is shown in orange and cloudy models (at the $f_{\rm sed}$ values of 0.01, 0.03, 0.1, and 0.3) in blue. The WFC3 data is shown in black points with error bars, and the black dotted lines correspond to the simulated phase curves from K15. As seen with the white-light phase curve, PICASO is able to closely reproduce the models presented in K15 for a cloud-free atmosphere, with exception at 1.49 $\mu$m (H$_2$O band center), when the peak of the two phase curves differ the most (∼100 ppm). We again attribute the discrepancy to differences in updated opacity and chemistry. The updated PICASO results are now closer to the observed data than those in K15. While the cloud-free models recover the peak in IR flux at all wavelength bins, the cloudy models provide the best overall fit to the data. As seen in the white-light curve, cloudy models particularly match the data well at phases after transit and before secondary eclipse.

To analyze the goodness of fit of the models, we compute the residual between our spectroscopic phase curves and the observations for each wavelength bin and orbital phase. Figure 7 shows residual heat maps for the cloud-free thermal flux ratio and the cloudy results with three different sedimentation efficiencies: 0.03, 0.1, and 0.3. We leave out $f_{\rm sed}=0.01$ from this analysis since it is very clear from Figures 4 and 6 that it does not provide a close fit to the observations. The residual heat map shows the absolute value of the difference between our PICASO models and the data, as a function of wavelength and orbital phase ($e=|r_{\rm obs}-r_{\rm model}|$). We easily see that the cloud-free phase curve is far from observations near the nightside, during and after transit, and before secondary eclipse (phases up to 0.3). This discrepancy is enhanced at wavelengths above 1.5 $\mu$m, as well as between 1.2 $\mu$m and 1.3 $\mu$m. All cloudy models are able to better reproduce observations near transit, but we find that higher values of $f_{\rm sed}$ provide better overall fits to the data, especially away from transit. We see that the model with $f_{\rm sed}=0.03$ is the worst fit away from transit for most wavelengths. Cloudy models with sedimentation efficiencies of 0.1 and 0.3 differ slightly from one another, with both fitting the nightside better than the cloud-free model while still fitting the dayside observations well. $f_{\rm sed} = 0.1$ is able to match nightside observations more closely, but we see little difference between the two cloudy models at phases very near transit. Overall, we conclude that a cloudy model with a sedimentation efficiency of around 0.3 provides the smoothest residual distribution and therefore the best match to observations. A more flexible model, with adjustable values of $f_{\rm sed}$, as well as an in-depth investigation into the cloud species that could be forming in the atmosphere of WASP-43b, will be helpful to further improve the model precision.

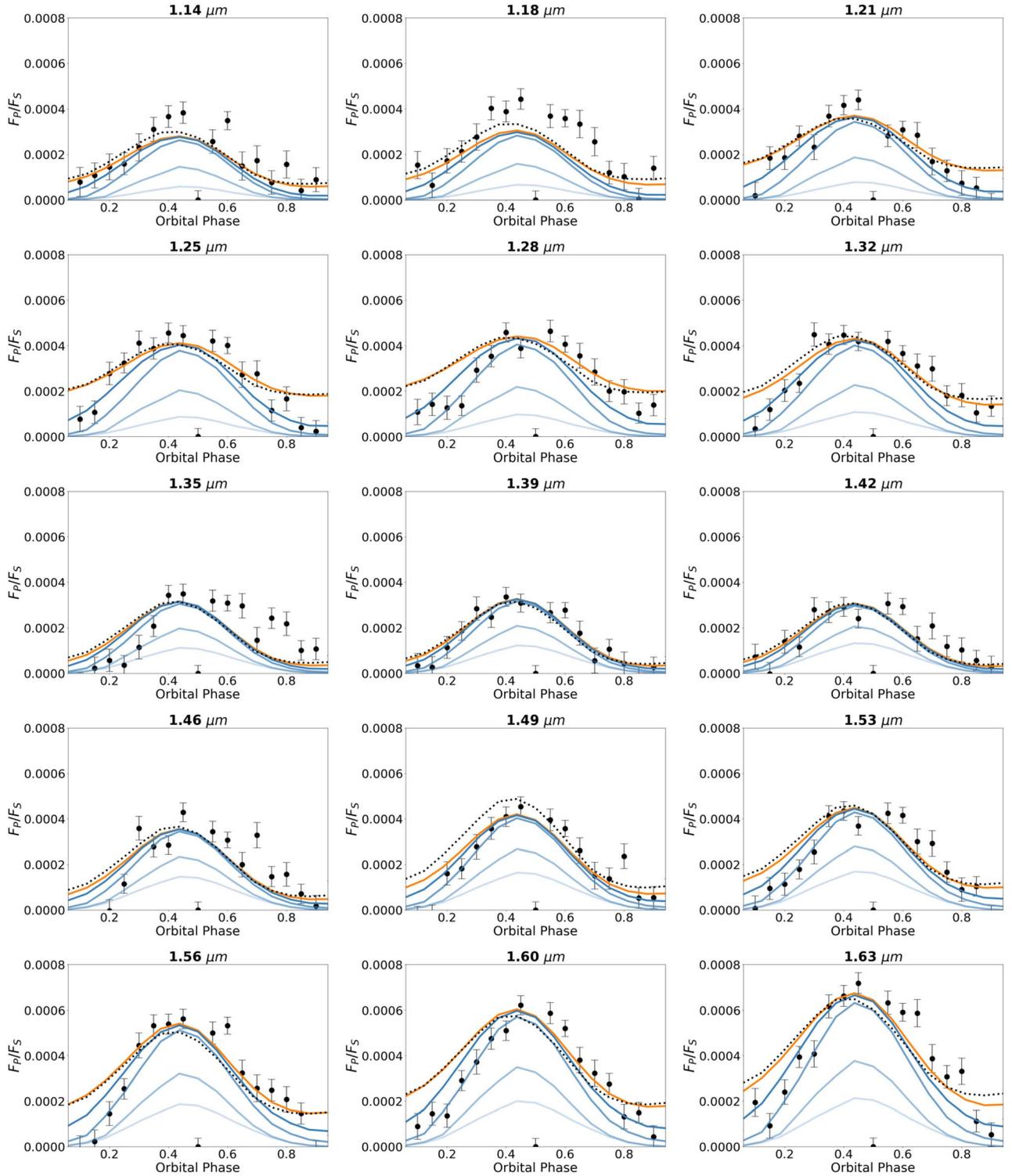

**Figure 6.** 1× solar-metallicity spectroscopic phase curves of WASP-43b 15 spectral bins from 1.125 to 1.650 $\mu$m, multiplied by the planet-to-star radius ratio $\left(\frac{R_P}{R_S}\right)^2$. Each bin consists of a 0.035 $\mu$m range. Our cloud-free model is shown in orange, and cloudy models with increasing $f_{\rm sed}$, corresponding to increasing line opacity, are shown in blue. Data is taken from WFC3 (Stevenson et al. 2014), and the black dotted line shows the 1× solar cloud-free model from K15.

For a more quantitative comparison of the models, we calculate the reduced chi-square ($\chi^2_{\rm red}$) corresponding to each model. These values are displayed in Figure 8. Low sedimentation efficiency (0.01 and 0.03) result in the highest $\chi^2_{red}$ values, and therefore these models represent the worst fit to the observations. Models with $f_{sed}$ values of 0.1 and 0.3 provide the optimal fit with respect to the cloud-free model, especially at longer wavelengths. This can also be seen clearly from the linear representation of the reduced chi-square values on the right panel of Figure 8, where most curves are minimized at $f_{sed}$ values of 0.1 and 0.3. In terms of wavelength, the best fit occurs at the band center for all models.

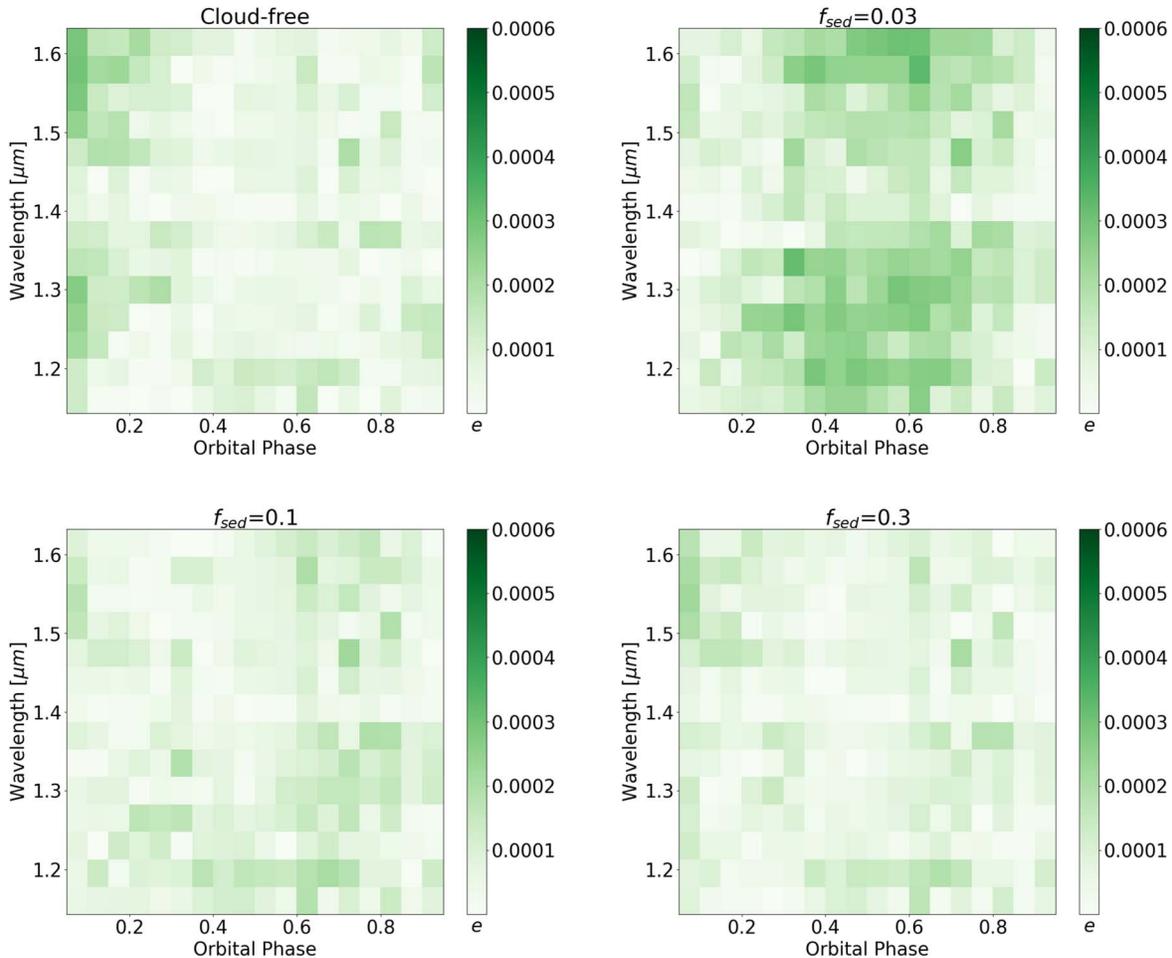

**Figure 7.** Residual heat maps for WASP-43b, showing the discrepancy between our PICASO models and the WFC3 data as a function of wavelength and orbital phase. The residual is calculated as $e = |r_{obs} - r_{model}|$ for every data point, and we compare cloud-free models with cloudy models using different values of $f_{sed}$. The residual at secondary eclipse was interpolated from the nearest values, since our model does not consider occultation of the planet by the star, but this event is visible on the WFC3 data.

We further compare our cloudy models with observations in Figure 9, which shows the emission spectra at all 15 observed phases, the phase-resolved planet-to-star flux ratio as a function of wavelength. Orange lines represent our PICASO cloud-free models, blue lines show our cloudy models with different sedimentation efficiencies, and dotted black lines are the 1× solar cloud-free models from K15. We are able to closely reproduce the K15 cloud-free models at all wavelengths and phases, with small discrepancies that we attribute to the different opacity/chemistry used between PICASO and K15 (from Fortney et al. 2008). We show that the dayside spectra (phases = 0.31–0.69), as well as phases after secondary eclipse and before transit (phases = 0.5–0.88), are better represented by the cloud-free model and the $f_{sed} = 0.3$ cloudy model. Again, we find that the cloudy models provide a better fit for the nightside and the phases after transit and before secondary eclipse (phases = 0.06–0.31). $f_{sed} = 0.3$ models deliver the spectra that most closely match the overall data. We find that including clouds has a much greater effect on the nightside emission than it does on the dayside, since a much thicker, higher altitude cloud deck is developed on the nightside. The upper cloud layer appears more opaque west of the substellar point, in agreement with predictions that clouds tend to form on the nightside, persist on the western edge of the dayside, and dissipate on the dayside.

It is important to note that K15 showed that a 5× solar cloud-free model provided a better fit to the data, particularly the white-light and spectroscopic phase curves, and especially the dayside emission. An enhancement in metallicity from 1× to 5× solar serves to increase the phase curve amplitude and reduce the peak in IR flux. However, K15 was unable to match the fluxes at nightside longitudes, due to the lack of clouds in their model. Even at solar metallicity, our PICASO cloudy thermal phase curves provide a much better fit to the nightside longitudes. It may be possible to provide a better fit to the WFC3 observations by including the cloud radiative feedback in the temperature calculation, incorporating microphysics to obtain the particle-size distributions (which will change the outgoing flux), or enhancing the metallicity of the model with the post-processed clouds.

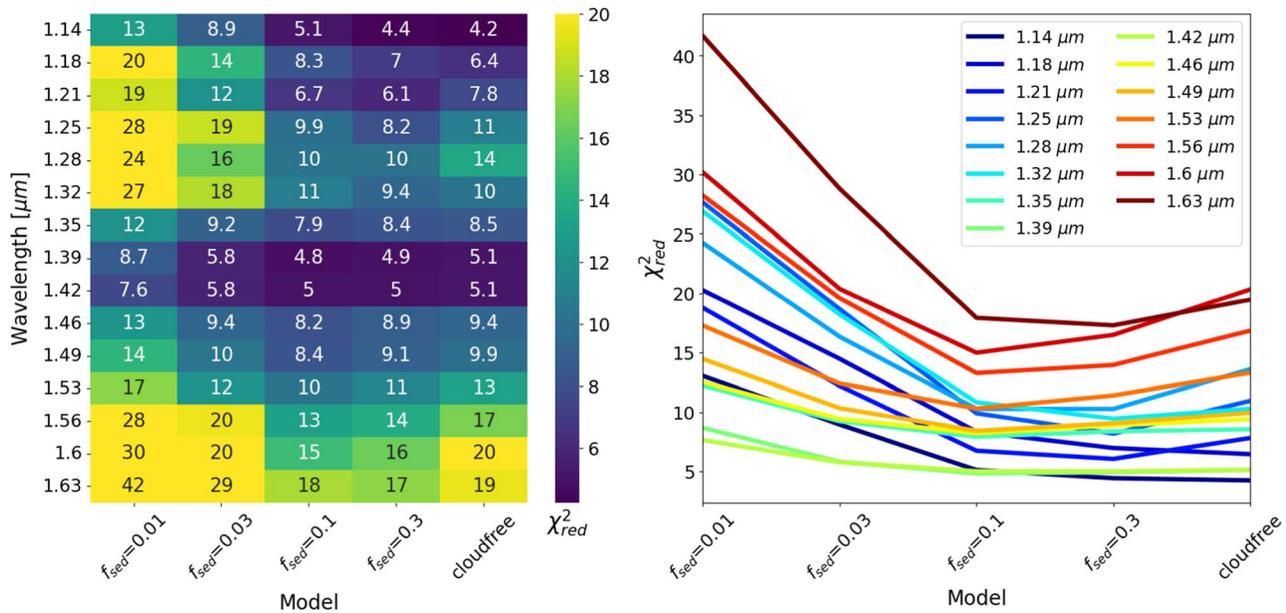

**Figure 8.** Left: heat map of reduced chi-square ($\chi^2_{\rm red}$) values for cloud-free and cloudy models, at all 15 wavelength bins, taking into account the uncertainty of the observations. Right: linear representation of $\chi^2_{\rm red}$, for every model and wavelength.

### 3.2.3. Thermal Phase Curves at 3.6 and 4.5 μm: Comparison to Spitzer/IRAC observations

We also compute thermal phase curves for WASP-43b at the Spitzer IRAC bands, 3.6 and 4.5 μm. There have been multiple observations and retrievals with this instrument for WASP-43b (e.g., Stevenson et al. 2017; Mendonca et al. 2018a; Morello et al. 2019; May & Stevenson 2020; Bell et al. 2021). We compare our models with the observations performed by Stevenson et al. (2017), which consist of two phase curves at 3.6 μm and one at 4.5 μm. The second visit at 3.6 μm and observations at 4.5 μm show day–night contrasts and phase curve peak offsets consistent with WFC3 data, which suggests that all three wavelength ranges probe similar depths in the atmosphere (Stevenson et al. 2017).

The reanalysis of the data performed by Mendonca et al. (2018a) brought the nightside into closer agreement with the cloud-free models (e.g., K15), but still predicted that a cloudy nightside was needed to match the observations. Using their open-source GCM, THOR (Mendonca et al. 2016), clouds were included in the GCM in the form of a constant, additional opacity covering the nightside of the planet. They conclude that dayside emission is matched both by cloud-free and cloudy models.

In Figure 10 we compare observations from Stevenson et al. (2017) with our cloudy ($f_{\rm sed} = 0.01, 0.03, 0.1,$ and 0.3) and cloud-free models, which were in turn compared to GCM phase curves from K15. We find that cloud-free models overpredict the nightside emission but provide a good fit for the planet dayside. All of our cloudy models match the nightside observations better, except for the first visit at 3.6 μm. This observed phase curve exhibits strong nightside emission that requires invoking nonphysical conditions in the retrieval models. A cloud-free patch on the hypothesized nightside cloud deck arising from variability could be a possible explanation, but it seems more likely that the first visit at 3.6 μm is a deceptive result (Stevenson et al. 2017; Morello et al. 2019). We consider the data at 4.5 μm to be more robust.

From this analysis, we show that the post-processed clouds computed with Virga have enough opacity to decrease the nightside emission at longer wavelengths. Figure 11 shows the chi-square values ($\chi^2$) for the cloudy and cloud-free models at 3.6 and 4.5 μm. Again, cloudy models provide a closer fit to the data according to the $\chi^2$. In this case, however, low $f_{\rm sed}$ values seem to provide a better overall fit, and especially at the nightside. One possible explanation to reconcile the difference between the best-fitting cloudy models for WFC3 and Spitzer data can be found in Mendonca et al. (2018a), where they suggest that an enhanced abundance of $CO_2$ on the planet's nightside could suppress the nightside flux. This was motivated by the ability of $CO_2$ to absorb radiation at 4.5 μm relative to other molecules. However, they found that this is unlikely in a later study using a more accurate chemical model (Mendonca et al. 2018b).

### 3.2.4. Limitations of Using Post-processed Clouds

In this work, we asynchronously couple PICASO and Virga, the latter of which computes post-processed clouds. While the GCM does contain radiative transfer for the molecular opacity of the atmosphere, the GCM is cloud-free and the radiative feedback from clouds is not considered. This is a limitation in the computation of our phase-resolved observations, as clouds have been shown to alter the thermal emission of planetary atmospheres in several ways (e.g., Roman & Rauscher 2019; Harada et al. 2021). Clouds can interact with outgoing thermal emission, thus altering the dynamics, chemistry, and thermal structure of an atmosphere through radiative feedback (e.g., Heng & Demory 2013; Lee et al. 2016, 2017; Lines et al. 2018; Roman & Rauscher 2019; Harada et al. 2021). Venot et al. (2020) investigated the radiative feedback of active MnS clouds on WASP-43b and found that, while reducing nightside flux, these clouds increased the dayside emission beyond what is detected in observations. Lines et al. (2019) self-consistently coupled EddySed (Ackerman & Marley 2001) to their GCM, and showed that the cloud opacity raises both the nightside and dayside flux, but especially the dayside, which leads to a higher day–night temperature contrast.

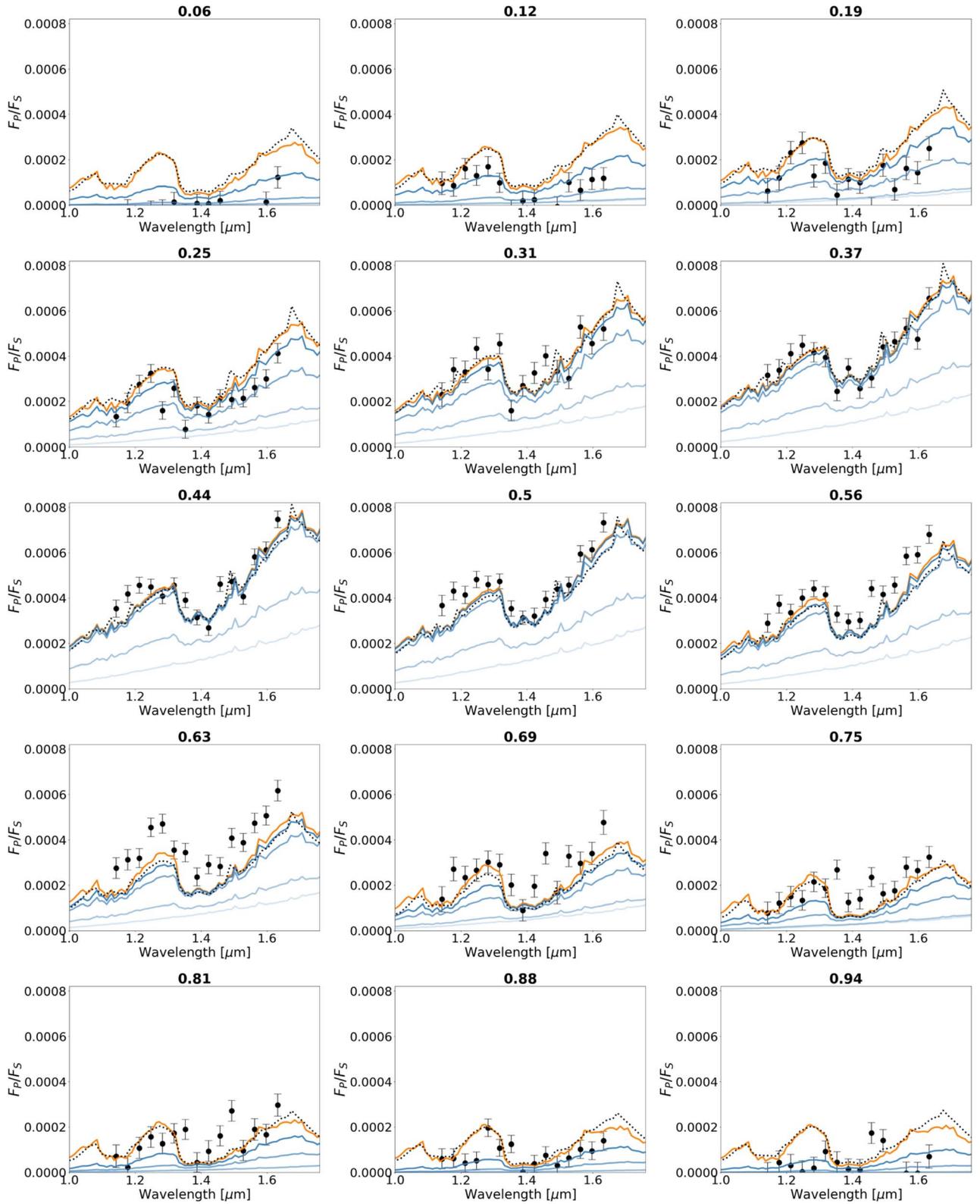

**Figure 9.** Phase-resolved thermal emission spectra for WASP-43b, specifically the planet-to-star flux ratio multiplied by the planet-to-star radius ratio $\left(\frac{R_P}{R_S}\right)^2$. The orbital phase angle is normalized, with the nightside visible at phases = 0.06, 0.12, 0.19, 0.81, 0.88, 0.94, limbs at phases = 0.25, 0.75, and dayside at phases = 0.31–0.69. We show the cloud-free model in orange and the cloudy models in blue. We chose four values for $f_{\rm sed}$: 0.01, 0.03, 0.1, and 0.3 (increasing value corresponds to increasing opacity of the lines). The 1× solar cloud-free results from K15 are shown with the black dotted line, and data is taken from WFC3 G141.

Further studies show that cloudy models increase the dayside emission at both short and long wavelengths for different $f_{\rm sed}$ values (Roman & Rauscher 2019; Christie et al. 2021).

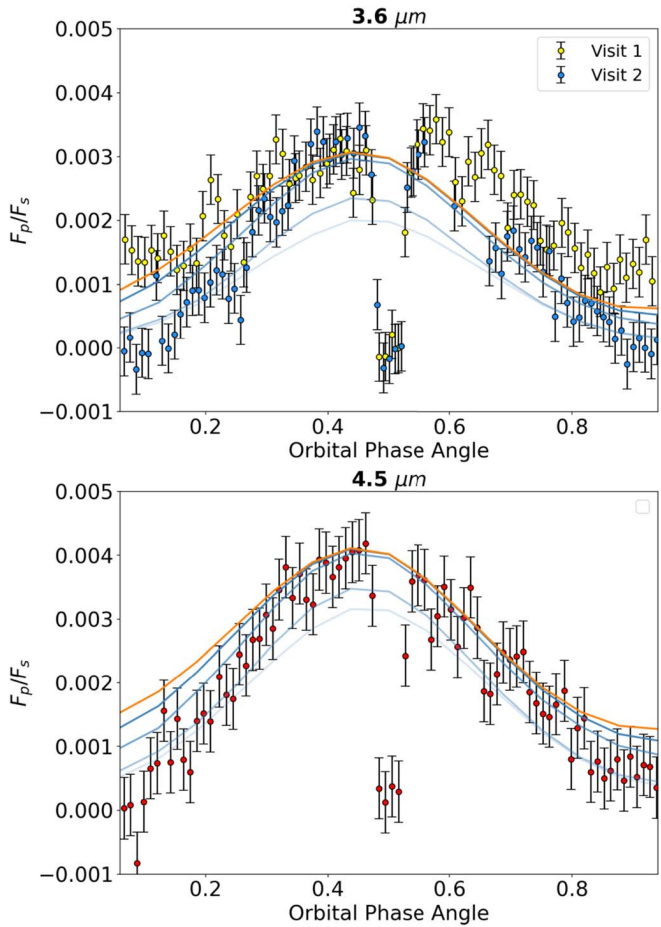

**Figure 10.** WASP-43b thermal phase curves at Spitzer/IRAC bands: 3.6 μm (top, blue, and yellow points) and 4.5 μm (bottom, red points). The data was obtained from Stevenson et al. (2017) and binned down to 120 points.

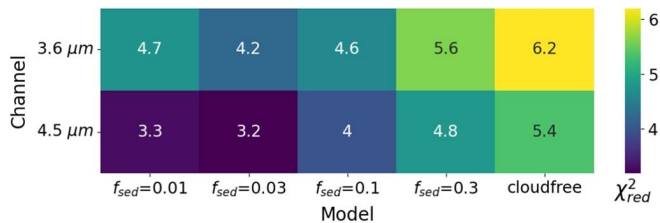

**Figure 11.** Reduced chi-square values for WASP-43b thermal phase curves at 3.6 μm (second visit) and 4.5 μm.

Parmentier et al. (2020) and Roman et al. (2021) found that reflective clouds on the dayside decreased global temperatures, while nightside clouds increased them.

While the formation and transport of clouds is certainly a coupled process between advection, radiation, and chemistry (Marley 2010), Venot et al. (2020) highlight how heavily dependent the amplitude and spatial distribution of cloud heating is on the model used (e.g., Lee et al. 2017; Roman & Rauscher 2017; Lines et al. 2018, 2019; Roman & Rauscher 2019; Roman et al. 2021). The use of post-processed clouds allows for greater flexibility and fast estimates of cloud properties, species, sedimentation parameters, and their effect on the thermal emission without the need to run computationally expensive coupled models (Venot et al. 2020; Adams et al. 2022). However, the lack of cloud radiative feedback might alter our conclusions about the best-fitting cloudy models. The presence of clouds contributes to the atmosphere's thermal structure, and this would have a strong effect on the dayside and nightside spectra. In our case, the clouds are mostly on the nightside, and thus even small values of $f_{\rm sed}$ (0.01 and 0.03, representing clouds that extend higher in the atmosphere with smaller particles), including cloud radiative feedback, could warm the whole atmosphere, increasing the dayside and nightside flux to provide a better match to the observations than shown here.

In addition, we only considered $f_{\rm sed}$ values up to 0.3, after confirming that results with higher values closely matched the cloud-free models. However, Christie et al. (2021) showed that even large values of $f_{\rm sed}$ can significantly alter the dayside flux by considering cloud radiative feedback with different mixing treatments. Thus, it is possible that $f_{\rm sed}$ values higher than 0.3 would still be able to decrease the nightside emission enough to better match the observations.

Another possible limitation of our approach could arise from the fact that we have included all species that are able to condense into clouds in Virga. A more detailed analysis of probable cloud condensates in the atmosphere of WASP-43b (accounting for, e.g., condensate surface energies) may improve our agreement with the data. In addition, we have used a constant value of sedimentation efficiency for the entire globe (all latitudes and longitudes, and all vertical layers). Modeling the 3D clouds with vertically dependent $f_{\rm sed}$ values in Virga could result in more realistic cloud structures (Rooney et al. 2022).

Observations of transiting exoplanets with the JWST will provide further insights to cloud formation and transport and can therefore provide further constraints to these theoretical models.

## 4. Conclusion

In this paper, we present a new formulation for computing thermal emission phase curves and spectra derived from 3D GCMs, built upon the framework of open-source codes PICASO and Virga. Until now, there have been no efficient methods within PICASO to compute 3D thermal spectra and phase curves from GCM output (Batalha et al. 2019). Because PICASO is coupled to Virga, this also allows the community to include the effects of clouds in the planet's atmosphere and a planet's outgoing thermal emission.

We apply this model to the hot Jupiter WASP-43b, which has been extensively observed and characterized with observations from the HST and Spitzer. We use the thermal and vertical wind structure obtained from the SPARC/MITgcm by K15 for a cloudless atmosphere, and post-processed them with PICASO assuming the presence of clouds using Virga. We then compute phase-resolved thermal emission in the form of white-light (band-integrated) and spectroscopic phase curves, and phase-resolved emission spectra. We compare our model results to previous simulated phase curves and spectra presented in Kataria et al. (2015, K15), as well as the HST/WFC3 observations from Stevenson et al. (2014) and Spitzer observations from Stevenson et al. (2017). We find good agreement between the 1× solar-metallicity cloud-free models presented in K15, thus validating our methodology. 5× solar-metallicity models, with and without clouds, may give a better match to the dayside and nightside data (K15); this investigation is left for future work. We also find that our PICASO-derived cloudy phase curves and spectra provide the best overall fit to WFC3 and Spitzer observations of WASP-43b. In our models, clouds have a greater effect on the nightside outgoing flux than on the dayside, since the cloud optical depth is larger on the cooler nightside. Including clouds into the atmospheric model allows for better representation of nightside emission, while fitting dayside observations nearly as well as cloud-free models. This is true for both WFC3 and Spitzer models. However, lower values of $f_{sed}$ provide a better match to Spitzer nightside data. It is important to keep in mind that the inclusion of cloud radiative feedback in the GCM model would likely warm up the atmosphere, particularly since these clouds are confined to the nightside, and would increase the thermal flux in the WFC3 and Spitzer bandpass. This could lead to a better fit for the lower $f_{sed}$ values than for the current best-fit value of 0.3 for WFC3. Nevertheless, the use of both post-processed and self-consistent clouds in 3D GCMs are complementary approaches that are useful in the understanding of clouds in exoplanet atmospheres.

Because the HST/WFC3 and Spitzer wavelength ranges are so narrow, we do not expect to see a large wavelength-dependent variation of the cloud opacity. This might become important, however, for future observations with the JWST, with instruments such as NIRISS, NIRCam, and MIRI, which span wavelength ranges of 0.6–2.5, 2.5–5, and 5–14 $\mu$m, respectively. Venot et al. (2020) showed that in simulated JWST/MIRI phase curve observations of WASP-43b, their retrieval models are able to distinguish between a cloudy and cloud-free nightside, but suggested that for clouds that are too thick or have a single scattering albedo in the IR that is too large, the models might struggle with the analysis of future JWST phase curve observations.

In the future, we can also convolve our PICASO phase curves with PandExo, a package for generating simulations for a suite of JWST and HST instruments (e.g., JWST/NIRSpec, NIRCam, NIRISS, and HST/WFC3) using through-put calculations from STScIs Exposure Time Calculator, Pandeia (Batalha et al. 2017). These simulated data could be used to make predictions for future JWST observations. Ultimately, this methodology presents a complementary approach to other open-source codes in computing phase-resolved emission derived from 3D models, while also including the effects of radiative transfer and clouds.

Upon acceptance of the manuscript, we plan to post the phase curve code to GitHub as a publicly available addition to PICASO.

We thank the anonymous reviewer for their helpful suggestions. This work was supported by the NASA Jet Propulsion Laboratory Education Office, Caltech Student-Faculty Programs, and ANRE Technologies Inc. through the JPL Summer Internship Program and the JPL Year-Round Internship Program. We thank Valerie Arriero for computing the 20 × 20 and 40 × 40 resolution phase curves. N.R.B thanks Xi Zhang for the advice and the useful discussions. N.B. acknowledges support from the NASA Astrophysics Division.

*Software:* picaso (Batalha et al. 2022), virga (Batalha et al. 2020a), numba (Lam et al. 2015), Matplotlib (Hunter 2007), pandas (McKinney 2010), pickle (Van Rossum 2020), bokeh (Bokeh Development Team 2014), NumPy (van der Walt 2011), SciPy (Virtanen et al. 2020), IPython (Perez and Granger 2007), Jupyter (Kluyver et al. 2016), PySynphot (STScI Development Team 2013), sqlite3 (sqlite3 Development Team 2019).

## Appendix
## Spatial Resolution Sensitivity Tests

We performed sensitivity tests with PICASO in order to determine which spatial resolution would allow for a balance between the computation time and precision of the model. We compare the cloud-free thermal phase curves computed with three different spatial resolutions: 10 × 10, 20 × 20, and 40 × 40 (Gauss angles × Chebyshev angles). These are shown in Figure 12. The three models yield close results, and the percentage error between one another does not surpass 3% for this cloud-free case. However, the three models vastly differ in the computation time and resources needed to run them, with the 10 × 10 model being the fastest and the least computationally expensive. The use of a high-performance computer was needed to run the 20 × 20 and 40 × 40 models.

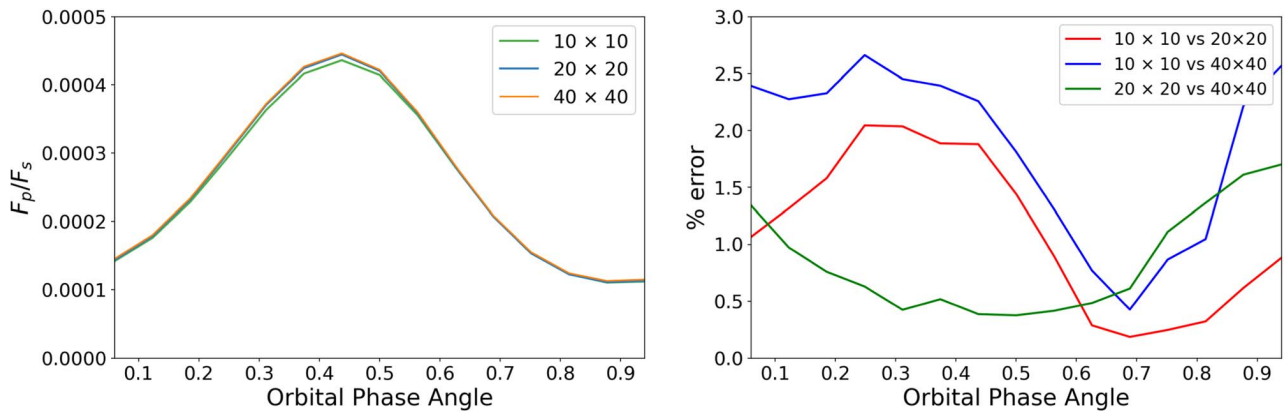

**Figure 12.** Left: comparison of three cloud-free thermal phase curves with different Gauss–Chebyshev spatial resolutions chosen in PICASO: 10 × 10, 20 × 20, and 40 × 40. Right: percentage error between each of the models, as a function of orbital phase.

ORCID iDs

Nina Robbins-Blanch https://orcid.org/0000-0002-8356-8712
Tiffany Kataria https://orcid.org/0000-0003-3759-9080
Natasha E. Batalha https://orcid.org/0000-0003-1240-6844
Danica J. Adams https://orcid.org/0000-0001-9897-9680